\documentclass[aps,prl,twocolumn,floatfix,reprint,square,comma,numbers]{revtex4-1}
\usepackage[latin9]{inputenc}
\usepackage{color,soul}
\usepackage{amsmath}
\usepackage{graphicx}
\usepackage{ulem}
\usepackage{amstext}
\usepackage{makeidx}
\usepackage{braket}
\usepackage[colorlinks=true,linkcolor=blue]{hyperref}

\begin{document}

\begin{@twocolumnfalse}
\title{Emulating anyonic fractional statistical behavior in a superconducting quantum circuit}

\author{Y. P. Zhong$^{1,2}$}
\author{D. Xu$^{1}$}
\author{P.~Wang$^{1}$, C.~Song$^{1}$, Q.~J.~Guo$^{1}$, W.~X.~Liu$^{1}$, K.~Xu$^{1}$, B.~X.~Xia$^2$}
\author{C.-Y. Lu$^{2,3}$}
\email{cylu@ustc.edu.cn}
\author{Siyuan Han$^{4}$}
\author{Jian-Wei Pan$^{2,3}$}
\email{pan@ustc.edu.cn}
\author{H. Wang$^{1,2}$}
\email{hhwang@zju.edu.cn}
\affiliation{
$^{1}$ Department of Physics, Zhejiang University, Hangzhou, Zhejiang 310027, China\\
$^{2}$ Synergetic Innovation Center of Quantum Information and Quantum
Physics, University of Science and Technology of China, Hefei, Anhui
230026, China\\
$^{3}$ CAS-Alibaba Quantum Computing Laboratory, Shanghai, 201315, China \\
$^{4}$ Department of Physics and Astronomy, University of Kansas,
Lawrence, Kansas 66045, USA
}

\date{\today}
\begin{abstract}
Anyons are exotic quasiparticles obeying fractional statistics,
whose behavior can be emulated in artificially designed spin systems.
Here we present an experimental emulation of creating anyonic excitations
in a superconducting circuit that consists of four qubits,
achieved by dynamically generating the ground and excited states of the toric code model,
i.e., four-qubit Greenberger-Horne-Zeilinger states.
The anyonic braiding is implemented via single-qubit rotations:
a phase shift of $\pi$ related to braiding, the hallmark of Abelian 1/2 anyons,
has been observed through a Ramsey-type interference measurement.
\end{abstract}
\maketitle
\end{@twocolumnfalse}

In three dimensions, elementary particles are classified as either fermions or bosons
according to their statistical behavior. In two dimensions,
the laws of physics permit the existence of anyons, which are exotic quasiparticles
obeying fractional statistics ranging continuously between the Fermi-Dirac and
Bose-Einstein statistics~\cite{Wilczek1982}. Although direct observation of anyonic excitations
and the associated fractional statistical behavior
in fractional quantum Hall system remains experimentally challenging,
artificially designed spin model systems may promise
an alternative and likely easier route
in light of certain theoretical treatments such as the toric code model~\cite{Kitaev2003,Kitaev2006}.

\begin{figure}[t]
  \centering
  \includegraphics[width=2.8in,clip=True]{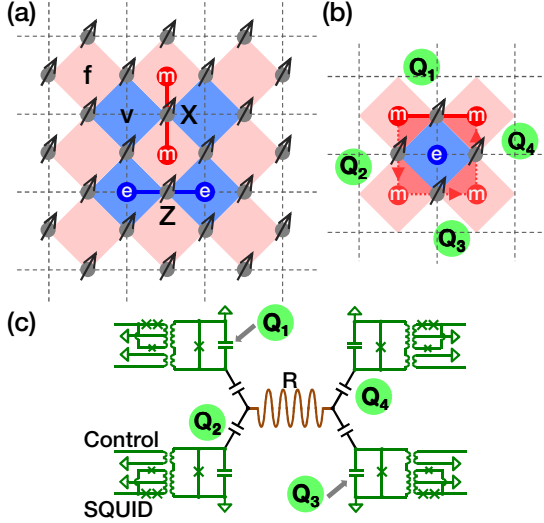}
  \caption{
	(a) Illustration of the toric code model. Qubits, symbolized by balls with arrows, are located on
	the edges of a two dimensional square lattice. The lattice is divided
	into two types of regions, the vertices (light blue) and the faces (light red),
	where $e$ particles and $m$ particles reside respectively. A pair of $e$
	particles ($m$ particles) can be created on neighboring vertices (faces)
	by applying a $Z$ ($X$) rotation on a qubit.
  (b) Four qubits, labeled from $Q_1$ to $Q_4$, of a vertex represent
	the minimal unit of the toric code.
	The panel illustrates the braiding action by moving an $m$ particle around an $e$ particle.
	(c) Schematic of the superconducting circuit featuring
	four qubits coupled to a central resonator. Arrangement
	of the four qubits correspond to those in (b). Also shown with each qubit
	are the control coil which tunes the qubit frequency
	and the integrated superconducting quantum interference device which
	probes the qubit state.
	}
	\label{fig1}
\end{figure}

The toric code is designed on a two-dimensional square lattice,
with qubits located on the edges (Fig.~\ref{fig1}(a)).
The model Hamiltonian is given by
\begin{equation}\label{toric}
  H = -\sum_{v}A_v - \sum_{f}B_f,
\end{equation}
where $A_v=\Pi_{j\in \text{star}(v)}X_{j}$ for each vertex $v$,
$B_f=\Pi_{j\in \text{boundary}(f)}Z_{j}$ for each face $f$,
and $X$ ($Z$) denotes the standard Pauli matrix $\sigma_x$ ($\sigma_z$).
$A_v$ and $B_f$ are called the stabilizer operators.
The ground state $|\psi_g\rangle$ of the Hamiltonian $H$
yields an eigenvalue $+1$ for both $A_v$ and
$B_f$ of all vertices and faces.
A quasiparticle called $e$ ($m$) particle is generated
on vertex $v$ (face $f$) if $A_{v}$ ($B_{f}$)
acting on the resulting state $|\psi_e\rangle$ ($|\psi_m\rangle$)
yields an eigenvalue $-1$.
A pair of $e$ ($m$) particles
are generated on the neighboring two vertices (faces)
by applying a $Z$ ($X$) rotation to qubit~$j$.
This can be understood according to the anticommutation relation of $Z$ and $X$,
i.e., $A_v(Z_j|\psi_g\rangle) = -Z_jA_v|\psi_g\rangle = -Z_j|\psi_g\rangle$
for the two vertices $v$s that connect to qubit $j$
and $B_f(X_j|\psi_g\rangle) = -X_jB_f|\psi_g\rangle = -X_j|\psi_g\rangle$
for the two faces $f$s that border qubit $j$.
Two particles of the same type on the same site
annihilate each other, i.e., the resulting state yields
the eigenvalue $+1$ for $A_v$ or $B_f$. The $e$ and $m$ particles
are anyonic excitations since their mutual statistics can be fractional.

It was observed in Ref.~\cite{Han2007} that the statistical properties of
anyons are associated with the underlying ground and excited states, which thus proposed that the anyonic fractional statistical behavior
can be studied by dynamically creating the ground and excited states of the toric code Hamiltonian.
This theory was previously only demonstrated with single photons and nuclear magnetic resonance
(NMR)~\cite{Lu2009,Pachos2009,Feng2013}. However, the liquid NMR system cannot prepare pure
quantum states and multipartite entanglement~\cite{Braunstein,Linden,Vidal}. The photonic experiments also suffered from an important drawback that the underlying ground state of the Hamiltonian, which were four-photon Greenberger-Horne-Zeilinger (GHZ) states \cite{Pachos2009} and six-photon graph states \cite{Lu2009}, were generated probabilistically with a low efficiency, and verified using post-selection, i.e., the photons had to be destroyed \cite{RMP}.

To remedy these problems, we turn to a solid-state physical system with the ability of deterministic preparation of the underlying entangled states and single-shot measurement of genuine multipartite entanglement. Here, utilizing a superconducting quantum circuit consisting of four phase qubits
coupled to a common resonator bus, we demonstrate anyonic fractional statistics
by deterministically creating four-qubit GHZ state and subsequently apply single-qubit rotations to realize braiding operations of the anyons.
Our experiment directly observe a non-trivial phase shift of $(0.983\pm0.007)\pi$
associated with anyonic braiding, unambiguously confirms
the fractional statistics of Abelian anyons.

Since anyonic excitations are perfectly localized quasiparticles in the toric code model,
a small-scale system is considered to be sufficient
for proof-of-principle demonstration of anyonic braiding statistics~\cite{Han2007}:
four qubits connected to a single vertex realize the minimum cell~\cite{Pachos2007}.
The Hamiltonian thus involves a vertex and four incomplete neighboring faces,
the latter of which are identified by the bordering links between qubits (Fig.~\ref{fig1}(b)):
\begin{equation}\label{toricReal}
  H = -A-B_1-B_2-B_3-B_4,
\end{equation}
where $A = X_1X_2X_3X_4$, $B_1=Z_1Z_2$, $B_2=Z_2Z_3$,
$B_3=Z_3Z_4$, and $B_4=Z_4Z_1$ (the subscripts label the qubits).
It can be shown that four-qubit GHZ state $|\psi_g\rangle=(|0000\rangle + |1111\rangle)/{\sqrt{2}}$
is the eigenstate of both $A$ and $B_j$ with the same eigenvalue $+1$,
and therefore $|\psi_g\rangle$ is the ground state of $H$ in Eq.~(\ref{toricReal}).
Starting from $|\psi_g\rangle$, an $e$ anyon can be created
at the vertex by applying a $Z$ rotation to one of the four qubits
(the paired anyon at the vertex outside of the four-qubit cell is ignored),
and the resulting excited state is $|\psi_e\rangle=(|0000\rangle - |1111\rangle)/{\sqrt{2}}$.
Similarly, by applying an $X$ rotation to one qubit,
a pair of $m$ anyons can be created on the neighboring two incomplete faces in Fig.~\ref{fig1}(b).

\begin{figure}[t]
  \centering
  \includegraphics[width=3.4in,clip=True]{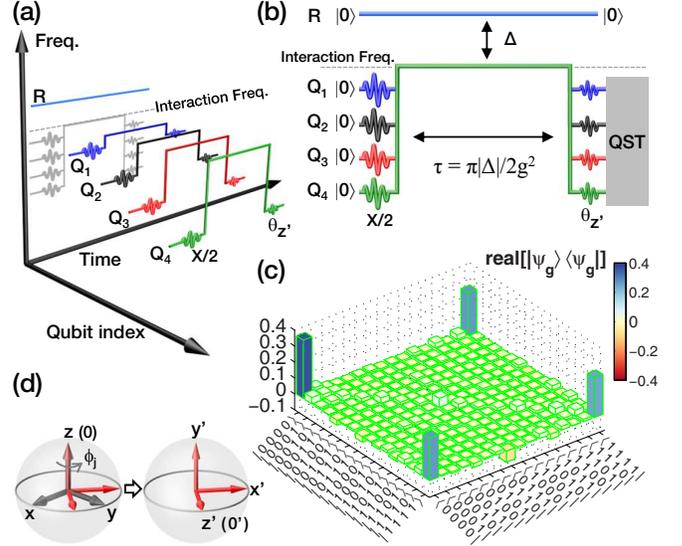}\\
  \caption{(a) Illustration of the pulse sequence in three dimensions
	for the one-step generation of the four-qubit GHZ state
	$|\psi_g\rangle$, where the three axes (frequency, time, and qubit index) are as labeled.
	For each color-coded qubit sequence line, the first 5 ns full width at half maximum (FWHM) Gaussian-shaped sinusoidal pulse
	realizes the $X/2$ gate~\cite{Lucero2008}, i.e., the $\pi/2$ rotation
	around the $x$-axis in the $x$-$y$-$z$ reference frame defined at the interaction frequency as labeled by the dashed line;
	the square pulse tunes the qubit to the interaction frequency;
	the second 4 ns FWHM sinusoidal pulse completes the phase-adjustment rotation
	$\theta_{z'}$ around the $z'$-axis in each qubit's own reference frame as defined in (d).	
	(b) The same pulse sequence projected onto the two-dimensional (2D) plane
	as defined by the frequency and time axes (see the shadow in (a)).
	QST is performed at the end to map out $|\psi_g\rangle$.
	(c) Real components of the density matrix $|\psi_g\rangle\langle\psi_g|$, where the prime sign
	on each $0'$ or $1'$ in the labels referring to the new frame is omitted for the clarity of the display.
	All imaginary components (data not shown) are measured to be no higher than 0.043.
  The polarization axis defining $|0'\rangle$ and $|1'\rangle$ is the $z'$-axis as illustrated
	in (d). The state fidelity $0.574\pm0.019$ exceeds the threshold of 0.5,
	confirming its genuine four-partite entanglement~\cite{Sackett,Guhne}.
	(d) Illustration of converting the reference frame $x$-$y$-$z$
	defined at the interaction frequency to the new $x'$-$y'$-$z'$ frame for qubit $j$ that picks up
	a dynamical phase $\phi_j$ in the pulse sequence.
	}\label{fig2}
\end{figure}

To realize the four-qubit minimum cell described by Eq.~(\ref{toricReal})
we use a circuit-quantum electrodynamics system with four qubits and
a resonator (see schematic in Fig.~\ref{fig1}(c)),
similar to that used in ref.~\cite{Lucero2012,Feng2015}.
The Hamiltonian is
\begin{equation}\label{hamiltonian}
  H_{\textrm{exp}}/\hbar = \omega_{r}a^{\dag}a + \sum_{j=1}^{4}{\omega_j}\sigma^{\dag}_j\sigma_j + g\sum_{j=1}^{4}(\sigma^{\dag}_{j}a + \sigma_{j}a^\dag),
\end{equation}
where $a$ ($a^{\dagger}$) is the lowering (raising) operator
of a single mode of the resonator, $\omega_r/2\pi=6.2$ GHz is its resonant frequency,
$\sigma_j$ ($\sigma^{\dag}_j$) is the lowering (raising) operator of qubit $j$,
$\omega_j/2\pi$ is the corresponding resonant frequency,
which is tunable from 5 to 7 GHz, and
$g/2\pi \approx 15.5$ MHz is the qubit-resonator coupling strength.
For each qubit, at its individual idle frequency where rotation pulses are applied (Fig.~\ref{fig2}),
the energy relaxation time $T_1\approx 600$ ns, and the dephasing time $T_{2}^{\ast} \approx180$ ns.
Here $T_{2}^{\ast }$ is obtained by fitting to
$\ln [P_{1}(\tau)]\propto -\tau /2T_{1}-(\tau /T_{2}^{\ast })^{2}$, where $P_1$ is the
$|1\rangle$-state probability of the Ramsey fringe envelope~\cite{Sank2012}.
However, in our numerical simulation
we find that the effective dephasing time $T_{2}^{\rm eff}$ in the Markovian master equation
has to be increased in order to explain our experimental results, likely due to the
following reasons: $T_{2}^{\ast}$ slightly increases as the qubit frequency increases;
for short pulse sequences the low frequency part of the noise spectrum has less impact on dephasing~\cite{Sank2012,Averin};
at the interaction frequency where excitations were effectively shared among all four qubits,
the impact by dephasing can be reduced due to cancellations of
the uncorrelated fluctuating noise environments for individuals,
which was investigated elsewhere~\cite{Averin}.

The four-qubit GHZ state is generated using the one-step protocol as
proposed in Ref.~\cite{Zheng2001} (Fig.~\ref{fig2}).
We first apply an $X/2$ (the ${\pi}/{2}$ rotation around the $x$-axis)
to each qubit at its idle frequency,
following which we apply a square pulse to bring the qubit to the interaction frequency,
where $\Delta/2\pi\equiv\omega_j/2\pi-\omega_r/2\pi \approx -57$ MHz.
The phase of each qubit's microwave
is calibrated according to the rotating frame at the interaction frequency~\cite{Neeley2010},
where the conventional $x$-$y$-$z$ coordinates associated with each qubit are defined.
Via the virtual photon exchange mediated by the resonator,
the qubits pick up dynamical phases that nonlinearly depend upon the collective qubit excitation numbers~\cite{Zheng2001}
and become maximally entangled after a duration of $\tau \approx \pi\Delta/2g^2$,
resulting in a GHZ state that is polarized along the $x$-axis with
$|\psi_0\rangle =  (\otimes_{j=1}^{4}|-\rangle_j-i\otimes_{j=1}^{4}|+\rangle_j)/{\sqrt{2}}$,
where $|-\rangle_j=(|0\rangle_j-|1\rangle_j)/\sqrt{2}$ and $|+\rangle_j=(|0\rangle_j+|1\rangle_j)/\sqrt{2}$.
At the end of the square pulse, each qubit returns to its idle frequency for further operations
and thus acquires a dynamical phase $\phi_j$, for qubit $j$, that is proportional to the product of the frequency change and the sequence time.
However, $|\psi_0\rangle$ can be formulated similarly as before if we define a new polarization axis for qubit $j$ that
rotates from the $x$-axis by an angle ${\phi_j}$ in the $x$-$y$ plane, i.e., we redefine
$|-\rangle_j=(|0\rangle_j-e^{i \phi_j}|1\rangle_j)/\sqrt{2}$
and $|+\rangle_j=(|0\rangle_j+e^{i \phi_j}|1\rangle_j)/\sqrt{2}$.

For convenience in analysis, here we change the reference frame as illustrated in Fig.~\ref{fig2}(d):
For each qubit, we use the new $x'$-$y'$-$z'$ Cartesian coordinates defined by the polarization axis (renamed as the $z'$-axis),
its perpendicular axis in the $x$-$y$ plane (renamed as the $x'$-axis),
and the $z$-axis (renamed as the $y'$-axis).
With $|0'\rangle$ and $|1'\rangle$ corresponding to the positive and negative
directions of the $z'$-axis, respectively, $|\psi_0\rangle$ can be simply rewritten as
$|\psi_0\rangle = (|0'0'0'0'\rangle + i|1'1'1'1'\rangle)/{\sqrt{2}}$.

In order to control the relative phase between $|0'0'0'0'\rangle$ and $|1'1'1'1'\rangle$ in $|\psi_0\rangle$,
we need to know the dynamical phase $\phi_j$, or equivalently the orientation
of the polarization $z'$-axis with respect to the $x$-axis for each qubit.
Aiming at maximizing the fidelity of the experimentally generated $|\psi_0\rangle$,
we perform the numerical optimization to locate the value of $\phi_j$,
based on which small phase-adjustment rotations ($\theta_{z'}$) around the $z'$-axis
are applied experimentally to accumulate a combined phase of $-\pi/2$, yielding the desired
ground state of the toric code model $|\psi_g\rangle = (|0'0'0'0'\rangle + |1'1'1'1'\rangle)/{\sqrt{2}}$.

To characterize $|\psi_g\rangle$ we perform the quantum state tomography (QST)~\cite{Neeley2008}.
The density matrix $\rho_g$ ($\equiv |\psi_g\rangle\langle\psi_g|$) of the experimentally generated
state $|\psi_g\rangle$ is shown in Fig.~\ref{fig2}(c), with a state fidelity of
$\text{Tr}(\rho_g\cdot\rho_g^{\text{ideal}})=0.574\pm0.019$.
This fidelity value confirms, with 3.9 standard deviations ($\sigma$), the genuine four-partite entanglement~\cite{Sackett,Guhne}.

\begin{figure}[t]
  \centering
  \includegraphics[width=2.2in,clip=True]{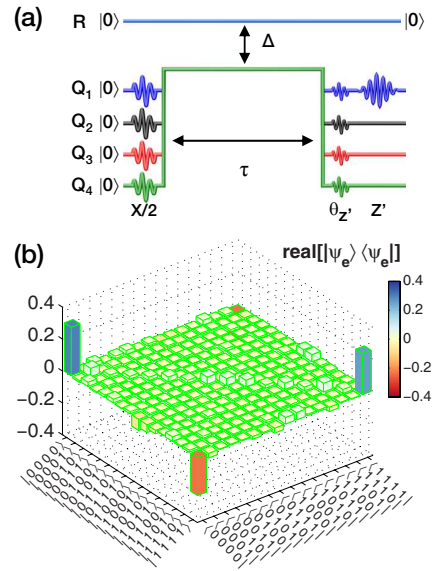}\\
  \caption{(a) Illustration of the pulse sequence in 2D
	for generation of the four-qubit GHZ state with the $e$-anyonic excitation $|\psi_e\rangle$.
	The sequence is similar to that shown in Fig.~\ref{fig2}(b),
	except that an extra 10 ns FWHM sinusoidal pulse is appended, completing the $Z'$ rotation on $Q_1$,
	to excite $|\psi_e\rangle$ out of $|\psi_g\rangle$.
	The QST pulses are not drawn.
  (b) The real components of $|\psi_e\rangle\langle\psi_e|$ generated by
	the pulse sequence in (a), where the prime sign
	on each $0'$ or $1'$ in the labels referring to the new frame is omitted for the clarity of the display. All imaginary components (data not shown) are
	measured to be no higher than 0.044. The state fidelity of $|\psi_e\rangle$ is $0.516\pm0.010$, indicating
	a genuine four-partite entanglement.}\label{fig3}
\end{figure}

Once the ground state $|\psi_g\rangle$ is prepared,
the excited state with an $e$ anyon at the vertex (Fig.~\ref{fig1}(b)),
$|\psi_e\rangle=(|0'0'0'0'\rangle - |1'1'1'1'\rangle)/{\sqrt{2}}$,
can be created by applying a $Z'$ rotation (a $\pi$ rotation around the $z'$-axis)
to one of the four qubits. For example, applying a $Z'$ on $Q_1$ (Fig.~\ref{fig3}(a)),
we obtain, by QST, the density matrix $\rho_e$ ($\equiv |\psi_e\rangle\langle\psi_e|$) with a state fidelity of
$0.516\pm0.010$ (Fig.~\ref{fig3}(b)),
which again confirms the genuine four-partite entanglement with $1.6\sigma$.

\begin{figure}[t]
  \centering
  \includegraphics[width=3.2in,clip=True]{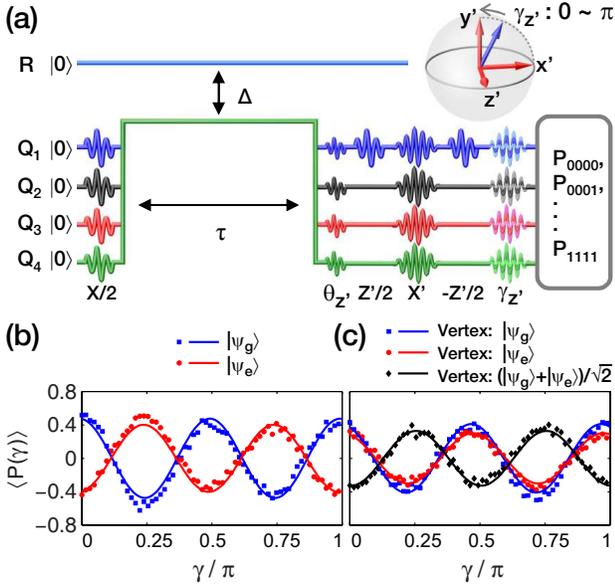}\\
  \caption{Correlation measurement demonstrating the anyonic braiding statistics.
	(a) The pulse sequence, for example, for measuring $P(\gamma)$ of braiding an $m$ particle around
	a half-filled vertex in the superposition state of $(|\psi_g\rangle+|\psi_e\rangle)/\sqrt{2}$:
	The first $Z'/2$ rotation following $\theta_{z'}$ creates the superposition;
	four simultaneous $X'$ rotations finish the braiding operation $C_{\textrm{loop}}$;
	the second $-Z'/2$ rotation returns the state to
	either $|\psi_g\rangle$ or $|\psi_e\rangle$ for the correlation measurement; the final
	10 ns FWHM sinusoidal pulse with varying amplitudes, $\gamma_{z'}$, rotates the state by an angle $\gamma$ ($0\sim\pi$)
	around the $z'$-axis, following which the four-qubit joint readout is performed, yielding the
	16 occupation probabilities, \{$P_{0000}$, $P_{0001}$, $\cdots$, $P_{1111}$\}. $\langle P(\gamma) \rangle$ is calculated
	as $\sum_{i_1,i_2,i_3,i_4}{(-1)^{i_1+i_2+i_3+i_4}P_{i_1,i_2,i_3,i_4}}$, where
	$i_j = 0$ or 1 refers to the state of qubit $j$.
	(b) Data of $P(\gamma)$ for the ground state $|\psi_g\rangle$ (the blue dots) and the $e$-anyonic excited state $|\psi_e\rangle$ (the red dots).
	The solid lines are fits according to the equation $P(\gamma) \propto \cos(4\gamma+\varphi)$,
	yielding $\varphi$s of $(0.033\pm0.008)\pi$ for $|\psi_g\rangle$
	and $(1.049\pm0.008)\pi$ for $|\psi_e\rangle$.
	(c) Data of $P(\gamma)$ for looping an $m$ particle around an empty vertex described by $|\psi_g\rangle$ (blue dots),
	an $e$-anyonic vertex described by $|\psi_e\rangle$ (the red dots),
	and a half-filled vertex in the state of $(|\psi_g\rangle+|\psi_e\rangle)/\sqrt{2}$ (black dots),
	with fits (the solid lines with corresponding colors) yielding $\varphi$s of $(0.135\pm0.007)\pi$, $(0.106\pm0.008)\pi$,
	and $(0.983\pm0.007)\pi$, respectively.
	}\label{fig4}
\end{figure}

In addition to QST, we also use the correlation measurement
to distinguish $|\psi_g\rangle$ and $|\psi_e\rangle$,
intended for a clear demonstration of anyonic braiding statistics later.
The distinction between the ground state $|\psi_g\rangle$ and
the $e$ anyonic excited state $|\psi_e\rangle$
is at the phase $\varphi$ between $|0'0'0'0'\rangle$ and $|1'1'1'1'\rangle$.
The correlation measurement allows for a direct probe of this phase by simultaneously measuring
four qubits along the same direction in the $x'$-$y'$ plane~\cite{Monz2011}.
Defining the correlation operator $P(\gamma)=\otimes_{j=1}^{4} (\cos{\gamma}Y'_j+\sin{\gamma}X'_j)$,
the expectation value of $P(\gamma)$ for GHZ state $(|0'0'0'0'\rangle +e^{i\varphi} |1'1'1'1'\rangle)/{\sqrt{2}}$
is $\langle P(\gamma) \rangle = \cos(4\gamma+\varphi)$, with a period of $\pi/2$ in $\gamma$
that is unique for four-qubit entanglement.
For phase qubit the polarization along the $y'$-axis ($z$-axis) can be measured directly.
Polarization along the axis of $\cos{\gamma}y'+\sin{\gamma}x'$
can therefore be measured after applying to each qubit a rotation by an angle
$\gamma$ around the $z'$-axis.
The measured $P(\gamma)$ versus $\gamma$ curves for $|\psi_g\rangle$ and
$|\psi_e\rangle$, both showing the Ramsey-type inference, reveal opposite phases as clearly visible in Fig.~\ref{fig4}(b).

Here we demonstrate the fractional statistics for Abelian 1/2 anyons, by braiding
an $m$ particle around an $e$ particle and
detecting the additional phase $\varphi$ due to braiding.
As shown in Fig.~\ref{fig1}(b), a pair of $m$ particles
can be created on the neighboring two incomplete faces
by applying an $X'$ rotation to $Q_1$.
Then one of them can be moved around the vertex
by successive application of $X'$ rotations to
the remaining qubits counterclockwise.
The two $m$ particles annihilate with
each other at last, completing the loop around
the vertex. This procedure can be described by
the loop operator $C_{\textrm{loop}}=X'_4 X'_3 X'_2 X'_1$.
Since the four Pauli operators in $C_{\textrm{loop}}$ commute
with each other, their exact ordering is not critical and
we simultaneously apply the four rotations in order to
minimize the impact of decoherence.
Looping an $m$ particle around an empty vertex gives no additional phase, i.e.,
$C_{\textrm{loop}}|\psi_g\rangle = |\psi_g\rangle$.
However, looping an $m$ particle around an $e$
particle yields a nontrivial statistical phase.
We first apply a $Z'$ rotation to generate an $e$
particle on the vertex, followed by
simultaneous $X'$ rotations to four qubits, fulfilling the loop
operation to circle an $m$ particle around the
$e$ particle. Finally we apply a $Z'$ rotation
again to annihilate the $e$ particle. A nontrivial
$\pi$ phase can be acquired, i.e.,
$Z'C_{\textrm{loop}}Z'|\psi_g\rangle = -|\psi_g\rangle$.

Although this additional phase $\pi$ does not
change the expectation values of
any Hermitian operator, it can be precisely retrieved by the
correlation measurement. To proceed, we generate the
superposition of the ground state and the $e$-anyonic excited
state by applying a $Z'/2$ operation to
$|\psi_g\rangle$, yielding $(|\psi_g\rangle-i|\psi_e\rangle)/\sqrt{2}$.
The loop operation $C_{\textrm{loop}}$ is applied subsequently, yielding
$(|\psi_g\rangle+i|\psi_e\rangle)/\sqrt{2}$.
The additional $\pi$ phase before $|\psi_e\rangle$
gained during the loop operation is crucial,
and a subsequent $-Z'/2$ rotation
brings the final state to $|\psi_e\rangle$ instead of $|\psi_g\rangle$,
which yields the opposite $\pi$ phase for the interference curve in the correlation measurement. As shown in Fig.~\ref{fig4}(c),
either looping an $m$ particle around an empty vertex
(the blue dots and line) or an $e$ particle (the red dots and line),
the correlation measurement yields $\varphi \approx 0$.
In contrast, the interference curve for looping an $m$ particle around a half-filled vertex in the superposition of $|\psi_g\rangle$ and $|\psi_e\rangle$ yields
$\varphi = (0.983\pm0.007)\pi$ (the black dots and line),
which clearly demonstrates the non-trivial phase shift of $\pi$
related to the braiding statistical of Abelian 1/2 anyons.

In conclusion, we have simulated the toric code model
for the first time in a solid-state quantum system.
A statistical $\pi$ phase related to the anyonic braiding has been observed.
With recent progress in superconducting quantum circuit technology~\cite{scalable},
our experimental methods could be further used to construct larger cluster states which can be used to demonstrate the robustness of topological braiding operations \cite{Han2007} and to explore topological quantum computation~\cite{topo}.

Note: A parallel experiment with ultracold atoms reported the realization of toric code and revealed anyonic statistics \cite{atom}.

\textbf{Acknowledgments.}
We thank J. M. Martinis and A. N. Cleland for providing the device used in the experiment.
This work was supported by the Chinese Academy of Sciences, NBRP of China (Grant Nos. 2012CB927404, 2014CB921201),
the NSFC (Grant Nos. 11222437, 11174248, 11434008), and the ZJNSF (Grant No. LR12A04001).
H.W. acknowledges support by the Fundamental Research Funds for the Central Universities of China (2016XZZX002-01). 
S.H. acknowledges support from NSF (Grant No. DMR-1314861).

\end{document}